\documentclass{jfm}
\usepackage{graphicx}
\usepackage{epstopdf, epsfig}
\usepackage{multirow}

\shorttitle{Solutocapillary flow induced in a waterbody by a solute source}
\shortauthor{I. Benouaguef, N. Musunuri, E. Amah , D. Blackmore, I. S. Fischer and P. Singh}

\title{Solutocapillary Marangoni flow induced in a waterbody by a solute source}

\author{Islam Benouaguef\aff{1}  ,
  Naga Musunuri\aff{1,3},
  Edison C. Amah\aff{1},
  Denis Blackmore\aff{2},
  Ian S. Fischer\aff{1}
 \and Pushpendra Singh\aff{1}
 \corresp{\email{singhp@njit.edu}}}

\affiliation{\aff{1}Department of Mechanical and Industrial Engineering, New Jersey Institute of Technology,
Newark, New Jersey, USA
\aff{2}Department of Mathematical Sciences, New Jersey Institute of Technology,\\
Newark, New Jersey,USA
\aff{3}Department of Mechanical Engineering, Indiana Institute of Technology, Fort Wayne, Indiana, USA}

\begin{document}

\maketitle
\begin{abstract}
The aim of this paper is to experimentally and analytically study the solutocapillary flow induced in a waterbody due to the presence of a solute source on its surface and the mixing induced by this flow of the solutes and gases dissolved at and near the surface into the waterbody. According to the analytic solution, the induced flow is analogous to a doublet flow in the sense that the flow is directed towards the source within a conical region with its vertex at the source, and outside the conical region the flow moves away from the source. The half cone angle for a negative source increases from $\sim$60 degrees with increasing source strength and Schmidt number reaching values greater than 80 degrees. When the cone angle is large, the outflow is restricted to a thin annular boundary layer region. These analytic results are in agreement with our experimental data obtained by the PIV (Particle Image Velocimetry) and PLIF (planar laser-induced fluorescence) techniques. As the solute gradient at the surface gives rise to the force that drives the flow, when the solute diffusion coefficient is reduced the flow becomes stronger and persists longer because the solute gradient is maintained for a longer time and distance. In experiments, the flow changes direction into the waterbody and the surface flow stops when the solute induced surface tension gradient driving the flow becomes comparable to the surface tension gradients that exist on the surface due to temperature gradients.
\end{abstract}

\begin{keywords}

\end{keywords}

\section{Introduction}
In this paper, we analytically and experimentally study the flow that arises in a waterbody due to the presence of a solute source on its surface. The flow arises because the solute source modifies the local surface solute concentration, and hence the surface tension of water, giving rise to a surface tension gradient which drives the flow \citep{Levich1969}. The surface tension gradient driven flows, which can arise also due to the surfactant concentration or temperature gradients, are important in a range of problems. They can cause enhanced mixing, e.g., rain drops falling on ocean surface, and are important in many industrial processes \citep{Grotberg1994,Dukhin1995,Ivanov1997,Eastoe2000,Eisner2005,Barnes2008}, e.g., in stabilizing emulsions and foams \citep{Carpenter1990, Craster2009}. These flows are also important in a range of biological processes such as in pulmonary flow where strain gradients induce surfactant concentration gradients \citep{Wood1993}, and some living organisms employ this mechanism for propulsion \citep{Harshey2003,Bush2006,Cantat2013}.

Although the focus of our work is on solutocapillary flow, the past studies conducted on the steady and transient surface flows caused by other surface tension altering sources present on a water surface, are also relevant to this study \citep{Jensen1995,Mizev2013,Roche2014,LeRoux2016,Mandre2017}. These studies note that on a clean water surface there is an axisymmetric radially outward flow on the surface which extends for a few centimeters from the source and then changes direction into the bulk as the surface flow becomes unstable due to periodic azimuthal perturbations. Beyond this radial distance, the surface flow has a multi-vortex structure.

The studies show that the distance for which the axisymmetric outward flow persists increases with increasing source strength. The flow also depends on the solubility of the surfactant used \citep{Bandi} and parameters such as the depth of the waterbody \citep{Jensen1995}. \citet{Roche2014} found the scaling laws for the radial distance over which the Marangoni flow persists \citep[also see][]{Napolitano1979, Suciu1967}. The water used in their study was Millipore water of high purity. Their data shows that the radius depends on the critical micelle concentration of the surfactant which was used to devise a method for measuring the critical micelle concentration of a surfactant from the strength of the flow induced on the water surface.

The strength of induced flow is significantly reduced when there are surface active contaminants which reduce the surface tension of water present in the waterbody \citep{Mizev2005}. For example, it has been shown that soon after water is poured into a container, or the water already in a container is mixed so that a fresh surface forms, the water surface remains relatively clean for a few minutes, especially when the concentration of surface active contaminants in the waterbody is small. Under these conditions, the radial distance over which the induced flow on the surface persists decreases with time as active contaminants continue to accumulate on the surface. After a few minutes the flow becomes unstable closer to source needle making the radius of the axisymmetric outward flow region smaller \citep{Mizev2005, Kovalchuk2006}. 
These studies also find that the radius over which the flow induced by a solute source persists is larger than that by a thermal source which the authors noted was due to the fact that it took them about two times longer time to setup a thermal experiment than one involving a solute. As discussed below, another plausible reason for this could be that the flow induced by a solute or surfactant is stronger because it takes the solute or the surfactant longer to diffuse away from the surface than heat.

  The steady state analytic solution of the thermocapillary problem was obtained by \citet{Bratukhin1967}. Specifically, they showed that for a positive heat source the flow on the surface is away from the source and below the source the liquid rises towards the surface, and for a negative source the opposite is true. \citet{Bratukhin1982} conducted a stability analysis of the axisymmetric flow which showed that it was unstable for very small source strengths leading to the formation of azimuthal vortices. This is consistent with a similar analysis done by Pshenichnikov \& Iatsenko (1974) for surfactant induced Marangoni flows. Also, the stability of the Marangoni flow depends strongly on the concentration of the surfactant, as the presence of a trace amount causes instabilities and ends the axisymmetric radial flow \citep{Mizev2013,Mizev2005, Kovalchuk2006, Bratukhin1982}.
 
 \citet{Bandi} have investigated the dependence of the surface water velocity on the distance from the source for water soluble and insoluble surfactants. Their data show that for both cases the velocity can be described using a power law model but with different decay coefficients. For an insoluble surfactant the radially outward surface velocity component decreases as $r^{-\frac{3}{5}}$, whereas for soluble surfactants it decreases as $r^{-1}$. In fact, when the Marangoni number is large enough a boundary layer develops near the surface this feature of the flow can be used to distinguish between these two types of surfactant behaviors. 
 
 The focus of most previous studies discussed above has been on the flow induced on a water surface and the flow below the surface has been described only qualitatively by noting that the water below the surface circulates back towards the source. \citet{Mandre2017} used numerical simulation results for soluble and insoluble surfactants to develop two distinct boundary layer theories to model the flows near the surface in the waterbody in terms of self-similar boundary layer profiles in the depth wise direction.
 
  In this paper, in addition to the flow at the surface, our focus is on the nature of flow in the waterbody and its variation with the source strength and other parameter values. Specifically, we present experimental and analytic results that show that the flow induced is analogous to that for a doublet (source-sink pair), with well defined in and out flow regions. For a negative solute source, the inflow occurs within a conical region and the outflow occurs outside the conical region. The half cone angle increases with increasing flow rate and so the outflow is restricted to a thin region near the water surface which as noted above can be described using the boundary layer approach. We also compare the analytic solution with the experimental results obtained using the PIV and the PLIF techniques. 

The key governing dimensionless parameters for the problem are the Reynolds number, $\Rey =\it u\hspace{1pt}  l/\nu$; Schmidt number  ${Sc = \nu /D }$ and the solutocapillary Marangoni number $Ma = ul/ D$. Here $l$ is the characteristic length, $u$ is the characteristic velocity, $\nu$ is the kinematic viscosity and $D$ is the salt diffusivity in water. Notice that there are no length and velocity scales present in this problem. Furthermore, the solute concentration at the source is not specified in the analytic solution, and only the solute flow rate is specified. The latter allows us to define a length scale $l= Q_c/{\mu}$ and a velocity scale is defined as $u = \Gamma / \mu$, where $\mu$ is the dynamic viscosity, $\Gamma = d\gamma/dc$, $\gamma$ is the surface tension, $c$ is the solute concentration, and $Q_c$ is the mass flow rate of solute. Using these length and velocity scales in the definition of Marangoni number, we have $Ma = (\Gamma Q_c)/ ({\mu}^2 D)$.
This expression for the Marangoni number agrees with a key dimensionless parameter that appears in the analytic solution described in section \ref{Analytic Solution}. For these length and velocity scales, $\Rey = Ma$ $Sc$. The diffusivity of salt $(NaCl)$ in water $D= 1.6\times10^{-9}$ $m^2/s$, $\Gamma= 0.0473\enskip mN/m.ppt$, and $Sc$ is equal to $447$. 

In the next section, we state the governing equations and obtain the analytic solution. This is followed by a discussion of the analytic and experimental results, and conclusions.

\section{Governing equations and boundary equations}
\begin{figure}
  \centerline{\includegraphics{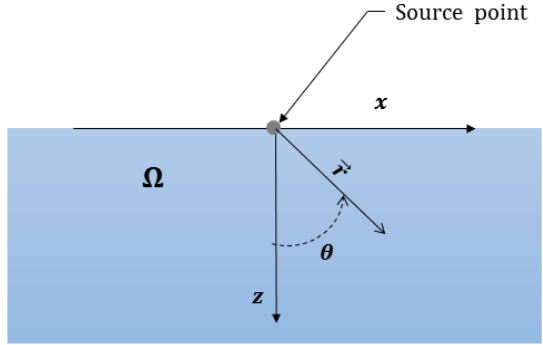}}
  \caption{ A source at the origin of a spherical coordinate system used to model the flow.}
\label{fig:sys_coor}
\end{figure}

We next state the equations governing the steady-state fluid velocity and concentration fields that result when a point solute source is present on the air-fluid interface of an incompressible fluid occupying the lower half-space $\{{\bf{\Omega}} := {x\in R^3 : 0 \le z,\enskip (0\le\theta \le \upi/2)}\}$.
We will use a spherical coordinate system to describe the flow with the source at its origin, as shown in figure \ref{fig:sys_coor}. 

The governing mass, momentum and solute concentration equations for the fluid system are:

\begin{equation}
    \nabla\bcdot\boldsymbol{u} = 0
    \label{cont_eq}
\end{equation}

\begin{equation}
    \rho(\frac{\partial \boldsymbol{u}}{\partial t} +\boldsymbol{u}\bcdot\nabla\boldsymbol{u} )
    = -\nabla p
    +\nabla\bcdot(2\mu\mathsfbi{D})
    +\gamma\kappa\delta({\bf\phi})\boldsymbol{n}
    -\delta(\phi) \nabla_s \gamma
    -\rho\boldsymbol{g}
    \label{moment_eq}
\end{equation}

\begin{equation}
    \frac{\partial c}{\partial t}+ \boldsymbol{u}\bcdot\bnabla c=
    D \nabla^2 c
    \label{advec_eq}
\end{equation}

Here $\rho$ is the fluid density, $\boldsymbol{u}$ is the velocity, $p$ is the pressure, $\mu$ is the dynamic viscosity, $D$ is the solute diffusion coefficient, $\mathsfbi{D}$ is the rate of strain tensor, $\gamma= \gamma(c)$ is the surface tension, $\kappa$ is the curvature of the air-fluid interface, $\delta(\phi)$ is the delta function, $\phi$ is the distance from the interface, $\boldsymbol{n}$ is the unit outer normal,$\nabla_s$ is the surface gradient, $c=(c_a-c_\infty)$ is the relative concentration, $c_a$ is the actual solute concentration, and $c_\infty$ is the solute concentration far away from the source which is assumed to be constant. Thus, far away from the source $c=0$.

The interface is assumed to be flat and characterized by $z=0$ or $\theta = \upi/2$ which implies $\kappa=0$. We will also assume that the fluid is incompressible with a constant viscosity, and the gravitational force can be neglected. The density of water decreases with decreasing the salt concentration. However, to keep our analysis simple, in this paper we will assume that the water density is constant.
The surface tension of water increases with the salt concentration, and so a change in the local surface salt concentration causes a surface tension gradient which induces a solutocapillary flow at the surface. We used experimentally measured data to obtain $\gamma=\gamma(c)$ from which we then obtained $\Gamma= {d\gamma}/{dc}$. Except for the source point, the boundary conditions on the interface$(z=0)$ are:

\begin{equation}
\nabla \bcdot(2\mu\mathsfbi{D}) = \nabla_s \gamma =  \frac{d\gamma}{dc}\nabla_s c   
\end{equation}

\begin{equation}
    \boldsymbol{u},c \rightarrow 0 \quad  as \quad  r \rightarrow \infty
\end{equation}

We will consider a steady-state solution of the equations. There is an obvious rotational symmetry about the $z$-axis, which means that in terms of spherical coordinates, the variables are independent of the azimuthal angle. Hence, $\boldsymbol{u}$ and $c$ are functions of $(r,\theta)$ with:
\begin{equation}
    \boldsymbol{u} = u_r(r,\theta)\boldsymbol{e_r}+u_\theta(r,\theta)\boldsymbol{e_\theta}
    \label{sep_var}
\end{equation}

The velocity and solute concentration in the domain are determined by the solution of the governing equations(\ref{cont_eq}-\ref{sep_var}). The problem is posed such that the mass flow rate of solute $Q_c$ at the source $(r=0)$ is specified, but the solute concentration itself is not specified. Since the solute flow rate is finite, the solute concentration at the source is singular.

For a negative solute source, the solute concentration decreases near the source causing a positive solute concentration gradient away from the source, i.e., the concentration increases with increasing distance from the source. For a positive solute source, on the other hand, there is a negative concentration gradient on the surface. The concentration gradient drives a solutocapillary flow away or toward the source. However, although there is a flow on the surface, there is no net liquid flow from (or into) the source itself.

In our experiments, the water flow rate from the source is not zero, as fresh or saltwater is injected at the surface using a syringe pump to modify the local salt concentration which is not the case for the theoretical model (equations \ref{cont_eq}-\ref{sep_var}). When freshwater is injected at the surface of a saltwater body, the salt concentration near the source decreases, and when saltwater is injected in a freshwater body, the salt concentration increases. The rate of inflow is held constant and so the effective solute inflow is also constant, as is the case for the analytic solution. Also, as discussed later, the fluid velocity due to the influx of water is negligible compared to that due to the solutocapillary flow. Therefore, although the experimental and analytic results deviate in the vicinity of the source, a short distance away from the source there is agreement between them.

\subsection{Analytic solution}
\label{Analytic Solution}
The governing equations \ref{cont_eq}-\ref{advec_eq} are mathematically identical to that for a point heat source present at air-water interface \citep[see][]{Bratukhin1967}, and so by using a similar approach, the velocity and concentration fields in spherical coordinates for equations (\ref{cont_eq}-\ref{sep_var}) can be written as:
\begin{equation}
    u_r(r,\theta)= \frac{\Theta_{r}}{r} = \frac{K\nu}{rR^2}
    \cos(\theta) RS + \sin^2{(\frac{\theta}{2}
    )
    [\alpha_1\xi^{2\alpha_2-1}+
    \alpha_2\xi^{2\alpha_1-1}+K]}
    \label{ur}
\end{equation}

\begin{equation}
u_\theta (r,\theta) = \frac{\Theta_\theta}{r} =
- \frac{K\nu\sin(\theta)S}{2rR}
\label{utheta}
\end{equation}

\begin{equation}
    c (r,\theta)=  \frac{\Theta_c}{r} =
    \frac{K\nu\mu(1+K)^{Sc}}{\Gamma r R^{2Sc}}
    \label{conc}
\end{equation}
\\
where :\\  
$\alpha_1:=\frac{1}{2}[1+\sqrt{1+K}], \enskip\alpha_2:=\frac{1}{2}[1-\sqrt{1+K}]$, 
$\xi= 1+\cos\theta$, $R=R(\theta)=\alpha_1\xi^{\alpha_2}-\alpha_2\xi^{\alpha_1}$, 
$S=S(\theta)=\xi^{\alpha_2-1}-\xi^{\alpha_1-1}$, 
$Sc =\nu/D$, and $\nu=\mu/\rho\quad$is the kinematic viscosity.

The dimensionless constant $K$ is determined by the solute flux condition:\\
\begin{equation}
    K(1+K)^{Sc} \int_0^{\frac{\upi}{2}}\{D+
    \frac{K\nu}{R^2}[(\sin^2{(\frac{\upi}{2})}+\cos\theta)T(\theta)+
    K\sin^2{(\frac{\upi}{2})}-\cos\theta]\}
    \frac{\sin\theta}{R^{2Sc}}d\theta =
    \frac{-\Gamma Q_c}{2\upi{\mu}^2}
    \label{solute_flux_condition}
\end{equation}

\begin{figure}
  \centerline{\includegraphics{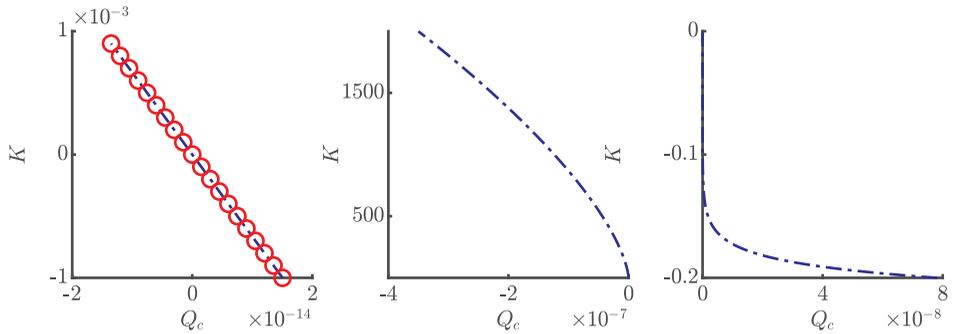}}
  \caption{$K$ as a function of $Q_c (kg/s)$ for $Sc = 447$. (a) $Q_c$ is small so that the value given by the analytic formula;oooo, and the numerically computed exact values;-.-.-., approximately agree. (b) $Q_c$ is negative. (c) $Q_c$ is positive.}
\label{fig:fig2_approx_num}
\end{figure}

An approximate value of $K$, which is appropriate when the solute in- or out-flow rate and Schmidt number are small, is $K=(-\Gamma Q_c)/(2\upi{\mu}^2 D) = Ma/(2\upi)$. But, when the flow rate and Schmidt number are not small, $K$ has to be obtained numerically by solving equation \ref{solute_flux_condition}. In figure \ref{fig:fig2_approx_num}, $K$ is plotted as function of the solute flow rate for a fixed value of Schmidt number $Sc=447$. The value of $K$ given by the approximate formula differs substantially from the exact value obtained numerically by solving the integral \ref{solute_flux_condition}. But, as figure \ref{fig:fig2_approx_num}a shows, it is reasonably accurate when $Q_c$ is very small. Figures \ref{fig:fig2_approx_num}(b) and \ref{fig:fig2_approx_num} (c) show the numerically computed values of $K$ for larger positive and negative values of $Q_c$. Notice that the magnitude of $K$ for a positive solute source is about three orders of magnitude smaller than for a negative source of the same strength and so the flow induced in this case is weak.

The computed value of $K$ can then be substituted in equations \ref{ur}-\ref{conc} to obtain the velocity and concentration fields. The separable analytic solution has the radial dependence of $1/r$ and so its complexity is primarily in the $\theta$ dependence of $\Theta_r$, $\Theta_\theta$ and $\Theta_c$ which depend on the parameter value (see equations \ref{ur}-\ref{conc}). A detailed discussion of the analytic solution is included in appendix \ref{app}.

The velocity and concentration fields on the surface, which take relatively simpler form, are given by: 
$c(r,\upi/2)= -(K\nu\mu)/(\Gamma r)$, 
$u_r(r,\upi/2)= K\nu/2r$, and $u_\theta(r,\upi/2)=0$. The surface tension gradient on the surface is 
$d\gamma/dr=(d\gamma/dc)\enskip  (dc/dr)
=\Gamma dc/dr= K\nu\mu /r^2$.
Notice that once the value of $K$ is known for the imposed $Q_c$, the velocity, concentration and surface tension gradient on the surface can be easily computed.

\subsection{Velocity doublet induced by a solute source and its dependence on $Sc$ and $Ma$}
\begin{figure}
  \centerline{\includegraphics{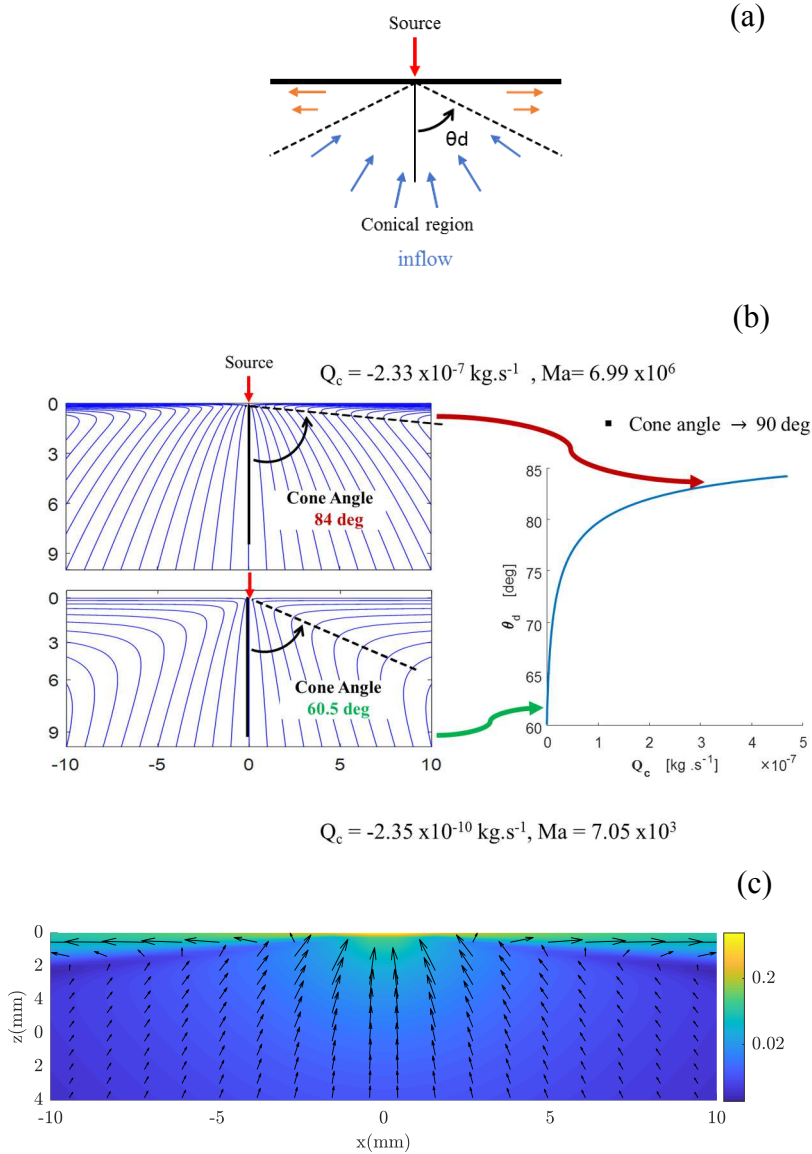}}
  \caption{(a) Schematic of the velocity doublet induced by a negative solute source. (b) The streamlines for two values of $Q_c$ and cone half angle $\theta_d$ as a function of $Q_c$. The half cone angle increases with $Q_c$. (c) Velocity field colored with velocity magnitude for a negative source with $K = 1525.8$, $Q_c = -2.33\times10^{-7} kg/s \enskip(Ma=7.0\times10^7)$. To improve clarity, the arrows near the origin (source) where the velocity is relatively large are not plotted, and the color bar is in the logarithmic scale.}
\label{fig:vel_doublet}
\end{figure}

As noted above, according to the boundary conditions prescribed at the source, the solute can enter or leave the source, but the inflow rate of liquid at the source is zero. The velocity and concentration fields given by equations \ref{ur}-\ref{conc}, satisfy these conditions. However, notice that although the net liquid inflow rate at the solute source is zero, $u_r$ and $u_\theta$ do not approach zero as $r$ approaches zero. Instead, they both approach infinity. The presence of a solute source on a liquid surface thus induces a flow doublet on the surface which causes flow in the half-plane occupied by the liquid. For a negative solute source the liquid below the source rises towards the source and on the surface it moves radially outward from the source, and both components increase in magnitude with decreasing $r$ (see figure \ref{fig:vel_doublet}). Thus, the source portion of the doublet is located near the upper surface and the sink portion near the vertical center line. The flow behavior for a positive solute source is qualitatively similar except that the flow direction is reversed.   
For a negative solute source $(Q_c< 0)$, $u_r$ is negative in a conical region with its vertex at the solute source and positive outside the conical region (see figure \ref{fig:vel_doublet}(c)). For a positive solute source, on the other hand, $u_r$ is positive in a conical region and negative outside the conical region (figure \ref{fig:vel_field_anal}). The cone angle depends on the system parameters including the solute flow rate and the solute diffusion coefficient (see figure \ref{fig:vel_doublet}(b)). Let us denote the cone half angle by $\theta_d $. For a negative solute source, $\theta_d$ approaches $60^{\circ}$ for small $Q_c$, and $\theta_d$ increases with increasing $Q_c$ to approach a value close to $90^{\circ}$ (see figure \ref{fig:vel_doublet}(b)). The outflow in the latter case is therefore restricted to a thin region near the liquid surface. For a positive solute source $(Q_c>0)$, for small $Q_c$, $\theta_d$ approaches $60^{\circ}$, and for larger $Q_c$, $\theta_d$ decreases with increasing flow rate to approach $59.5^{\circ}$. The maximum change in the cone half angle with the solute flow rate for the latter case is only about $0.5^{\circ}$, and so the streamlines are relatively rigid. For a negative solute source, on the other hand, the maximum change in the cone half angle is about $30^{\circ}$.

\begin{figure}
  \centerline{\includegraphics{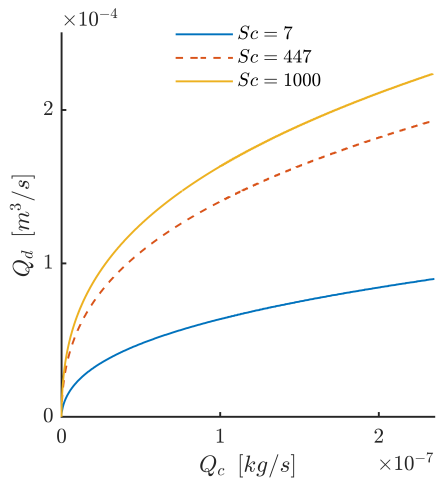}}
  \caption{The solutocapillary volume flow rate $Q_d$ for a negative solute source, as a function of $Q_c$ for $r =1$.}
\label{fig:qd_sink}
\end{figure}

The volume flow rate of liquid in the conical region (towards or away from the solute source) at any distance $r$  from the solute source is equal to the volume flow rate outside the conical region (away or towards the solute source) which ensures that the net liquid inflow from the solute source is zero. The volume flow rate in the conical region however decreases with decreasing $r$. These facts can be used to quantify the doublet strength by computing the flow rate, $Q_d$, towards or away from the source at a unit radial distance $(r = 1)$ from the source in the conical region (see figure \ref{fig:qd_sink}). The flow rate quantifies the volume of liquid moved per unit time by the solutocapillary flow, and can be used to compare the doublet strength for different parameter values. Notice that for a given $Q_c$, the flow rate $Q_d$ increases with increasing $Sc$ (or decreasing $D$, as $Sc$ was increased by decreasing $D$ while holding all other parameters constant). This is due to the fact that when $D$ is reduced the solute remains on the surface for a longer duration of time, and thus the surface tension gradient and the flow it causes are stronger.

\begin{figure}
  \centerline{\includegraphics{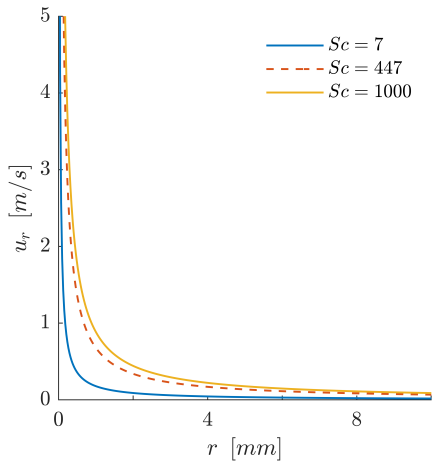}}
  \caption{The surface radial velocity as a function of r is shown for three values of Sc, obtained by reducing $D$ while keeping all other parameters fixed for a solute flux $Q_c=2.33\times10^{-7}\enskip kg/s$.}
\label{fig:ur_surface_r}
\end{figure}

The liquid velocity at the surface also increases with decreasing solute diffusion coefficient $D$ or increasing Schmidt number while all other parameter values are held fixed. The data presented in table \ref{tab:table_1} shows that $K$ increases when $Sc$ is increased, and from equations \ref{ur} and \ref{utheta} we know that the velocity increases with increasing $K$. Consequently, as figure \ref{fig:ur_surface_r} shows, the velocity on the surface increases with increasing $Sc$. For example, if $Sc$ is increased from $447$ to $1000$ (say, by using a solute with smaller diffusion coefficient), the velocity on the surface at a distance of $1 \enskip cm$ from the source increased by a factor of $1.3$.

The distance at which the velocity on the surface becomes smaller than a given fixed value also increases with increasing $Sc$. Since the diffusion coefficient is inversely related to the molecular weight, this suggests that a solute with a larger molecular weight will cause a stronger solutocapillary flow than that by a solute with a smaller molecular weight when their molar concentrations are the same. This is consistent with the experimental results reported in \citet{Roche2014} for the soluble surfactants of a given type which show that the flow caused by the surfactant with larger molecular weight persists over a longer distance (However, the diffusion coefficients of the surfactants used in the study were not provided.). The reason for this behavior reported in \citet{Roche2014} and \citet{LeRoux2016} is given in terms of the critical micelle  concentration (cmc) which for the surfactants considered decreases with increasing molecular weight. 

\begin{table}
    \centering
    \begin{tabular}{ p{1.2cm} p{1.2cm} p{1.2cm} p{1.2cm} p{1.2cm} p{1.2cm} p{1.2cm} p{1.2cm} p{1.2cm}  }
 \hline
    
    \multirow{1}{*}{}& 
    \multicolumn{4}{p{4cm}}{$Q_c=2.33\times10^{-8} [kg/s]$} &
    \multicolumn{4}{p{4cm}}{$Q_c=2.33\times10^{-7} [kg/s]$}\\
    \\

 $Sc$   & $K$  &$\alpha_c[deg] $ & $\alpha_m[deg]$ & $\delta_m/\delta_c$ & $K$  &$\alpha_c[deg] $ & $\alpha_m[deg]$ & $\delta_m/\delta_c$\\
 
 \hline
 $7$    &  $89$  & $11.1$  &$18.3$   &$1.7$    &$402$    &$5.0$ &$10.7$ &$2.1$\\
 $447$  &  $329$ & $0.6$   &$11.5$   &$17.9$   &$1524$   &$0.3$ &$6.2$ &$21.0$\\
 $1000$ &  $430$ & $0.3$   &$10.4$   &$27.7$   &$1992$   &$0.2$ &$5.7$ &$32.5$\\
 \hline
\end{tabular}

    \caption{For selected values of $Q_c$ and $Sc$, the values $K$, $\alpha_m$, $\alpha_c$ and $\delta_m/\delta_c$ are shown for the analytic solution}
    \label{tab:table_1}
\end{table}

\subsection{Solute and momentum boundary layers}
As discussed above, the solute remains concentrated at the surface for a longer distance from the source with increasing $Sc$, and so the solute concentration decreases sharply in the direction normal to the liquid surface. To quantify the latter, we may define the solute boundary layer thickness $\delta_c$ to be the distance from the surface at which the concentration decreases to one percent of the concentration at the surface. Let us consider points $(r,\pi/2)$ and $(r/\sin{\theta},\theta)$, on the normal to surface. The solute concentrations at these points are: 
$-K\nu\mu /(\Gamma r)$ and $-K\nu\mu(1+K)^{Sc}\sin{\theta}/(\Gamma r R^{2Sc})$, respectively (see equation \ref{conc}). Using the condition that the latter concentration is $0.01$ times the concentration at the surface, we obtain the following equation:
\begin{equation}
    0.01(\frac{K\nu\mu}{\Gamma})= \frac{K\nu\mu{(1+K)}^{Sc}\sin{\theta}}{\Gamma \enskip R^{2Sc}}
\end{equation}

The above is a transcendental equation which can be solved for the angle $\theta =\upi/2-\alpha_c$, where the solute concentration reduces to $1\%$ of the concentration on the surface. Here $\alpha_c$ is a small constant angle which depends on $Sc$ and $Q_c$, and other parameter values appearing in the governing equations. Notice that the angle does not depend on $r$. The equation can be solved numerically. In the spherical coordinate system, the equation for the concentration boundary layer is $\theta=\upi/2-\alpha_c$. The concentration boundary layer thickness at distance $r$ then is $\delta_c=r\sin{\alpha_c}$. 

Similarly, we may define the momentum boundary layer thickness $\delta_m$to be the distance at which the velocity magnitude decreases to one percent of the magnitude at the surface. The momentum boundary layer equation is given by $\theta=\upi/2-\alpha_m$, where $\alpha_m$ is a small constant angle, and the thickness at distance $r$ is $\delta_m=r\sin{\alpha_m}$.

Both, $\delta_m$ and $\delta_c$ increase linearly with $r$ due to the separable form for the analytic solution. Thus, $\delta_m/\delta_c$ is independent of $r$. In \citet{Mandre2017}, a boundary layer approximation was developed to model the tangential velocity profile for soluble and insoluble surfactants near the surface by assuming that the flow has axisymmetric cylindrical symmetry and is self-similar which was observed in the numerical results reported in the paper. It was noted that this boundary layer approximation, which is valid at large $Ma$ and away from the source, did not satisfy the Marangoni stress condition at the interface. The velocity profile given by equations \ref{cont_eq}-\ref{advec_eq}  being the exact solution satisfies the governing equations including the stress boundary conditions on the surface.

In table \ref{tab:table_1} the solute and momentum boundary layers are described for typical parameter values. For example, for $Sc = 447$ and $Q_c= 2.33\times10^{-7} kg/s$ $(K = 1524.4)$, the solute boundary layer is given by $\theta=89.7^{\circ}$ and the momentum boundary layer by $\theta=83.75^{\circ}$. So for $r=1 cm$,  $\delta_m=r\sin{6.25} = 1.09 mm$ and $\delta_c =r\sin{0.3}= 0.05 mm$. (We have chosen $r=1 cm$ for describing our results because a typical solutocapillary flow persists for only a few centimeters.) Thus, $\delta_m/\delta_c\approx20.97$. 
Notice that the thicknesses of both momentum and solute boundary layers is small, which as discussed below, makes the measurement of the velocity and solute profiles within the boundary layer challenging.

On the other hand, if it is assumed that $Sc = 7$ and $Q_c=2.33\times{10}^{-7} kg/s$, the solute boundary layer is given by $\theta=84.96^{\circ}$ and the momentum boundary layer by $\theta=79.44^{\circ}$. Thus, for $r=1 cm$, $\delta_m=r\sin{10.66} = 1.85 mm$ and $\delta_c =rsin{(5.04)}= 0.88 mm$, and $\delta_m/\delta_c\approx2.15$. Notice that the thickness of the solute boundary layer in this case is about $18$ times larger than for the case with $Sc = 447$, and the momentum boundary layer is about $1.7$ times thicker. These examples show that when the Schmidt number is increased both $\delta_c$ and $\delta_m$ decrease and, as a result, the surface flow persists for a longer distance. {\it{This is a consequence of the fact that solute is not a passive quantity in this problem. In fact, its concentration gradient at the surface gives rise to the force that drives the flow and so when D is reduced, the flow becomes stronger and persists longer because the solute gradient on the surface is maintained for a longer time and distance}}.

The above case is interesting also because the Schmidt number in this case is comparable to the Prandtl number $(\Pran=\nu/\kappa )$ for the thermocapillary flows, where $\nu$ is the kinematic viscosity and $\kappa$ is the thermal diffusivity. For water $\Pran\approx7$ (which is about 64 times smaller than the solute Schmidt number). We remind the reader that the thermocapillary flow induced by a heat source is governed by mathematically similar equations and the analytic solution has a similar form, except that the Prandtl number, which is analogous to the Schmidt number, is $\approx7$. Thus, when all other parameters are the same, the solutocapillary flow is expected to remain strong over a longer distance than the corresponding thermocapillary flow as the solute concentration gradient would persist for a longer distance. Furthermore, as discussed below, for the thermocapillary case the doublet half angle is smaller and so the inflow towards the source is at a steeper angle. 

\section{Experimental setup and results}

\begin{figure}
  \centerline{\includegraphics{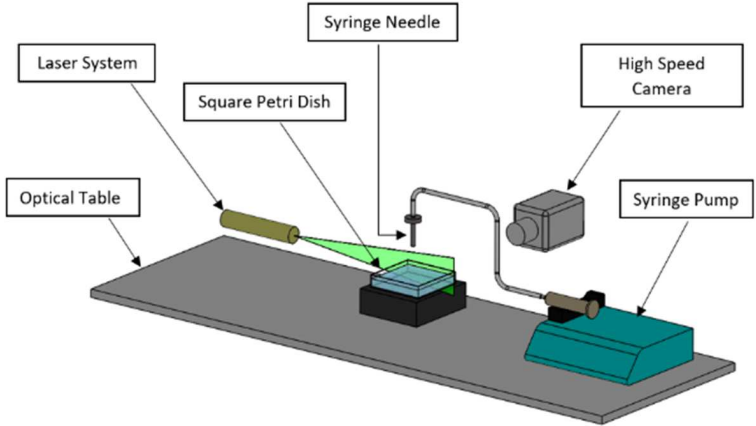}}
  \caption{Schematic of the experimental setup used for create a point freshwater source on the surface of a salt waterbody}
\label{fig:exp_setup}
\end{figure}

The experimental apparatus consisted of a glass aquarium with a square cross-section which is partially filled with water (Millipore water, resistivity, $18.2 M\Omega.cm$ ) (see figure \ref{fig:exp_setup}). A solutocapillary flow is generated by releasing fresh or salt water onto the surface by using a needle held perpendicular to the surface touching the water surface. The flow rate from the needle was controlled by a syringe pump. The transient solutocapillary flow in the waterbody was measured in a vertical plane (normal to the camera axis) illuminated by a laser sheet by the PIV (Particle Image Velocimetry) technique. The cross-section and depth of the container were varied to ensure that its size was large enough so that the lateral flow on the surface was not constrained by the container side walls. The vertical position of the camera was in line with the water surface, providing an undistorted view of the volume directly below the water surface. A high-speed camera was used to record the motion of seeding particles visible in the laser sheet. Then an open-source code, PIVlab, was used for the time-resolved PIV analysis. PIVlab is a Matlab based software which analyses a time sequence of frames to give the velocity distribution for each frame. A MatLab code was used for post-processing and plotting results \citep{Thielicke2014}.

We also employed the planar laser-induced fluorescence (PLIF) technique in which $CO_2$ is used as a flow tracer and its transport allows us to visualize the solutocapillary flow. A $1.0\times 10^{-5} mol/l$ fluorescein sodium (Uranine $C_{20}H_{10}Na_2O_5$) solution was used to monitor the concentration of dissolved $CO_2$ in the water body on a vertical plane illuminated by a laser sheet. At this concentration, the surface tension of water remained approximately equal to that of pure water \citep{Takagaki2014}. The fluorescence intensity of the Uranine solution decreases with decreasing pH, and thus since the pH decreases with increasing $CO_2$ concentration, the areas with increased $CO_2$  concentration appear darker. 
In our experiments, pressurized $CO_2$ from a cylinder was released via a diffuser into a closed chamber above the water surface increasing the concentration of $CO_2$ in the chamber and absorption of additional $CO_2$ into the waterbody. The newly absorbed $CO_2$ was concentrated in a thin layer near the surface. The layer appeared darker in the PLIF experiments as its fluorescence intensity was weaker. Dissolved $CO_2$ slowly diffused into the waterbody. The solutocapillary flow convected the $CO_2$ rich water near the surface in the flow direction with the latter serving as a flow tracer.

In our experiments, the local surface salt concentration of the waterbody was reduced by injecting fresh water at a constant volume flow rate onto the surface. This approach is necessary because it is difficult, if not impossible, to remove salt locally from the water surface at a fixed rate. The volume of fresh water injected in our experiments was much smaller than the volume of the waterbody and so the change in the average salt concentration of the waterbody due to the addition of water was small. The inside diameter of the syringe needle was $0.2 mm$ which ensured that the water was injected onto a relatively small area on the surface. 

Obviously, when the concentration of salt in the source water is equal to that in the waterbody a solutocapillary flow is not induced since the addition of water does not change the surface salt concentration. But, when freshwater is injected onto the surface the salt concentration near the source decreases which causes a solutocapillary flow (see movie 1). Notice that since the salt concentration is reduced near the source, this case is equivalent to the case when a negative salt source is present on the surface. In order to quantify the reduction in the salt concentration, notice that the addition of freshwater causes a salt deficit equal to the volume of freshwater added times the salt concentration in the waterbody, i.e.,
\begin{equation}
    Salt\enskip deficit = c_\infty V_s
    \label{salt_deficit}
\end{equation}
where $V_s=Q\Delta t$ is the volume of freshwater added in the time interval $\Delta t$, $c_\infty$ is the solute concentration of the waterbody before the addition of freshwater and $Q$ is the volume flow rate from the source. The average concentration after the addition of freshwater is $c_{av}=c_\infty V_b/ (V_b+V_s) \approx c_\infty,$ 
where $V_b$  is the volume of the waterbody and the volume of freshwater added is small,$V_b >>V_s$. 
An equivalent negative solute flow rate $Q_c$  can be obtained by dividing the salt deficit given by equation \ref{salt_deficit} by the time interval $\Delta t$ 
\begin{equation}
    Q_c= -\frac{c_\infty V_s}{\Delta t}= -c_\infty Q
    \label{qsolute}
\end{equation} 			

From this equation we notice that a constant freshwater injection rate ensures an effective constant salt outflow rate. This relation also allows us to quantitatively compare our experimental measurements with the theoretical results presented in section 3. 
Also note that the volume flow rate of water from the needle, which was varied between $50-400 \mu l/min$, made negligible contribution to the overall flow rate. In fact, when the solute concentration of the injected water is equal to that of the waterbody, and so there is no induced solutocapillary flow, the flow a short distance from the needle was barely observable.

\begin{figure}
  \centerline{\includegraphics{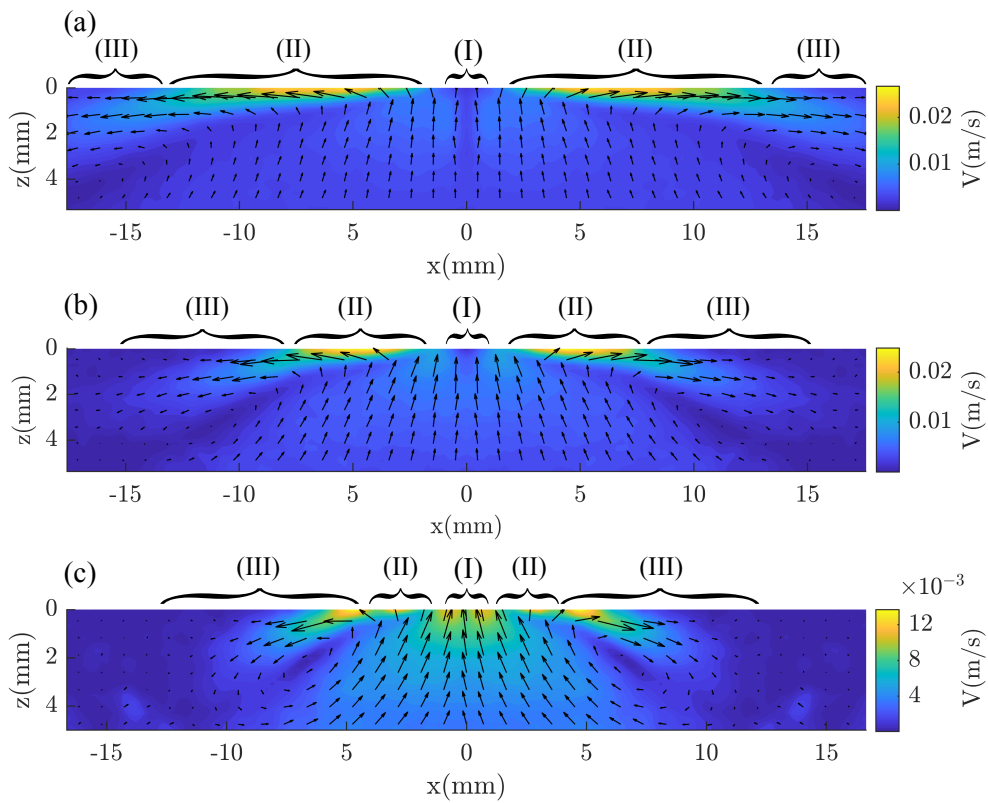}}
  \caption{The velocity field on a vertical plane passing through the center of the needle used to inject fresh water onto the surface obtained using the PIV technique. The distribution on the right side of the needle was measured and the on the left side is the mirror image of the former. The solute concentration in the waterbody was $35 ppt$. (a) $Q_c= -2.33\times 10^{-7} kg/s$ ($K=1525.8$). (b) $Q_c = -1.17\times 10^{-7} kg/s$ ($K=961$) and (c) $Q_c= -0.58\times 10^{-7} kg/s$ ($K=606$). Region (I): Flow influenced by the needle. Region (II): Flow matches the analytic solution. Region (III): Surface flow changes direction inwards.}
\label{fig:piv_vel_3fr}
\end{figure}

In figure \ref{fig:piv_vel_3fr}, the velocity fields for three flow rates are shown on a vertical plane passing through the center of the syringe needle used to inject fresh water onto the surface of the saltwater body. Notice that the velocity field obtained using the PIV technique is qualitatively similar to that for the analytic solution presented in figure \ref{fig:vel_doublet}(c) for a negative solute source, except near the source needle and after a few centimeters away from the needle where the surface flow changes direction into the waterbody. The inflow is in a conical region and the outflow is outside this conical region. As is the case for the analytic solution shown in figure \ref{fig:vel_doublet}(c), the velocity magnitude along the cone boundary is locally minimal and the cone angle increases with increasing freshwater flow rate. The half cone angle for the induced flow doublet for $50 \mu l/min$ is $71.5^{\circ}$ and for $100 \mu l/min$ is $82^{\circ}$. The corresponding values for the analytic solution are $79.1^{\circ}$ and $84.5^{\circ}$, respectively. The size of the region in which the flow is strong increases with increasing freshwater flow rate.

Also, as previous experimental studies have noted, the solutocapillary flow weakens with increasing distance from the source and at a distance of a few centimeters from the needle depending on the volume flow rate from the needle, its direction turns into the waterbody \citep{Mizev2013,Roche2014,Mandre2017}. This, as discussed below, is accompanied by the formation of vortices which cause gases dissolved near the surface to mix in the bulk water. The distance for which the surface flow persisted increased with increasing flow rate from the needle (see figure \ref{fig:piv_vel_3fr}). For the flow rates of $Q_c= -2.33\times 10^{-7} kg/s$, $Q_c = -1.17\times10^{-7} kg/s$ and $-5.83\times 10^{-8}  kg/s$, the distances were $2.5$, $1.7$ and $0.9\enskip cm$, respectively.

\begin{figure}
  \centerline{\includegraphics{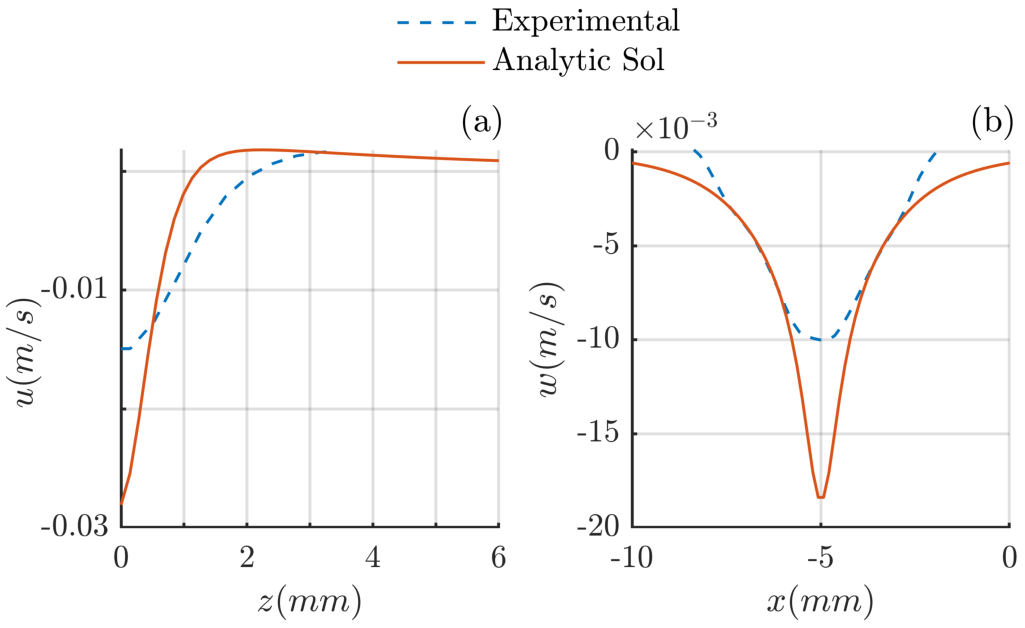}}
  \caption{Velocity distribution on a vertical plane passing through a negative source for $Q_c = -5.83\times 10^{-8} kg/s$ ($K = 606$). (a) $u$ as a function of $z$ for $x = 6.14 mm$. (b) $w$ as a function of $x$ for $z = 2.0 mm$.}
\label{fig:exp_anal_50}
\end{figure}

\begin{figure}
  \centerline{\includegraphics{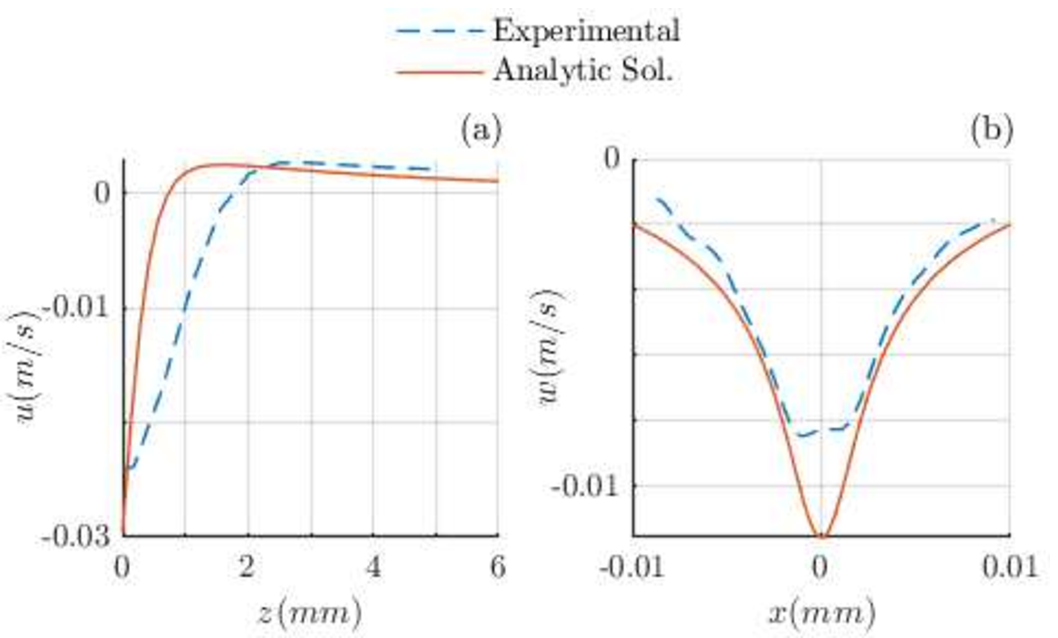}}
  \caption{Velocity distribution on a vertical plane passing through a negative source for $Q_c = -5.83\times 10^{-8} kg/s$ ($K = 606$). (a) $u$ as a function of $z$ for $x = 6.14 mm$. (b) $w$ as a function of $x$ for $z = 2.0 mm$.}
\label{fig:exp_anal_100}
\end{figure}

In figure \ref{fig:exp_anal_50}(a) the $x$-component of velocity obtained from the PIV data is plotted as a function of $z$ for a flow rate of $50 \mu l/min$ and in figure \ref{fig:exp_anal_50}(b) the $z$-component of velocity is shown as a function of $x$. In figure \ref{fig:exp_anal_100} similar data is shown for $Q = 100 \mu l/min$. These figures also show the corresponding velocity components for the analytic solution which are obtained by first computing $Q_c$ using equation \ref{qsolute} and then substituting it in equation \ref{solute_flux_condition} to compute the value of $K$. In figures \ref{fig:exp_anal_50} and \ref{fig:exp_anal_100}, the values of $K$ are $318$ and $606$, respectively. These values of $K$ were then used in equations \ref{ur} and \ref{utheta} to obtain the velocity for the analytic solution. There is good agreement between the PIV data and the analytic solution considering that the value of $K$ is determined using the solute flux condition, and so there are {\it{no}} adjustable parameters in the analytic solution. 

The momentum boundary layer thickness for the flow in figure \ref{fig:exp_anal_50}(a) is around $2 mm$ and for the analytic solution around $1.5 mm$, and in figure \ref{fig:exp_anal_100}(a) they are $0.8 mm$ and $1.3 mm$, respectively. Notice that the measured values of $u$ at and near the surface are smaller than the analytic values and the opposite is the case slightly below the surface, as the high gradients in the boundary layers near the surface were not accurately resolved in the PIV measurements. This is due to the reduced accuracy of the PIV technique near the water surface as the velocity is obtained by averaging over a volume with width that is not much smaller than the momentum boundary layer thickness which as noted earlier is about $1 mm$. Moreover, the averaging volumes at and near the surface are centered below the intended points. Similarly, from figures \ref{fig:exp_anal_50}(b) and figure \ref{fig:exp_anal_100}(b), we note that the measured values of $w$ and the analytic solution are in good agreement, except near $x = 0$, where the velocity for the analytic case varies sharply. The experimental values are smoother again because of the finite size of the averaging volume.

Notice that according to the analytic solution, the flow on the surface slows down with increasing distance from the source, but its direction does not change into the bulk. We postulate that in experiments the surface flow changes direction because of the presence of other naturally occurring adverse surface tension gradients which interfere with the flow. For example, the water surface is cooler than the bulk because of the evaporative cooling which gives rise to Rayleigh-Bernard convection cells that are accompanied by temperature and surface tension gradients on the surface \citep{sapngenberg1961}. Notice that these convection cells due to evaporative cooling begin to form within $70 s$.

We next estimate the distance after which the magnitude of solute induced surface tension gradient becomes comparable to that for the Rayleigh-Bernard (RB) convection. The RB surface tension gradients arise because of the temperature variation, and thus $(d\gamma/dr)_{RB}=  (d\gamma/dT) \enskip  (dT/dr)$. For water, $d\gamma/dT=1.7\times10^{-4}   N/(m^{\circ}C)$, \citep{sapngenberg1961} and so $(d\gamma/dr)_{RB}= 1.7\times10^{-4}\enskip dT/dr$. Setting the solute surface tension gradient $d\gamma/dr=K\nu\mu/r^2$   equal to that for the RB convection, we have $d\gamma/dr=K\nu\mu/r^2   = (d\gamma/dr)_{RB}= 1.7\times10^{-4}   dT/dr$. The distance at which the two  gradients are equal, $r_c=\sqrt{(K\nu\mu)/(1.7\times 10^{-4}  dT/dr)} = 76.7\sqrt{(K\nu\mu)/(dT/dr)}$.

From the experimental and numerical simulation data of the RB convection due to evaporation under standard conditions, $dT/dr\approx 0.1 ^{\circ}C/1 cm  =10 ^{\circ}C/m$ (estimate from Fig. 10 in \citet{sapngenberg1961}). Using this in the expression above, $r_c  = 76.7\sqrt{K\nu\mu/10}= 24.3 \sqrt{K\nu\mu}$. Thus, for  $K=1525.8$, which corresponds to the case shown in Fig. 15a, $r_c= 2.9 cm$. This distance is of the same order as the distance for which the surface flow persisted in figure \ref{fig:piv_vel_3fr}(a). The distances for $Q_c = -1.17\times 10^{-7} kg/s$ $(K=961)$ and $-5.83\times 10^{-8}  kg/s$ $(K=606)$, are $2.4$ and $1.89$ cm, respectively. This model for the critical distance, with no adjustable parameters, thus approciamately estimates the distances measured in our experiments.

We also observed that soon after the water was vigorously mixed the solutocapillary flow on the surface was stronger and changed direction into the waterbody at a longer distance. However, after about a minute, which is the time interval after which the RB convection cells form on the surface, the flow slowed and changed direction at the same distance as before. This, however, does not rule out the possibility that there are other mechanisms that interfere with the flow and come into play after a similar time interval. For example, the presence of a trace amount of surfactant, which may also accumulate at the surface after a similar time interval, can also diminish the strength of solutocapillary flow. This possibility was however ruled out for our experiments as the experiments were performed using ultra-pure water.

\subsection{PLIF visualization and solute transport due to solutocapillary flow}

\begin{figure}
  \centerline{\includegraphics{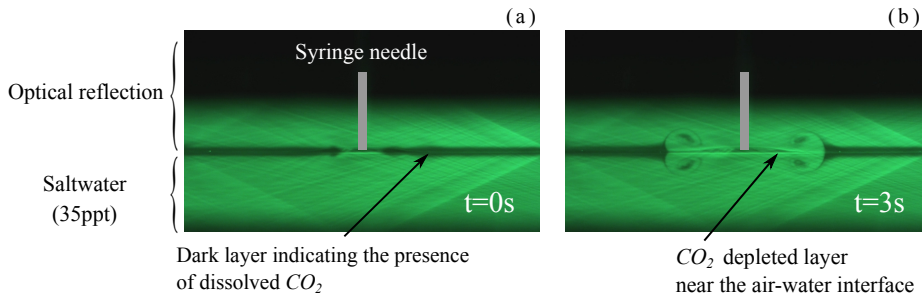}}
  \caption{PLIF images of $CO_2$ concentration profiles as it is being transported by the flow induced by a solute source present in the middle. (a) Shortly after $CO_2$ was injected in the chamber it is concentrated near the surface in the black band. The actual thickness of the layer is one half of the band thickness as the upper half of the band is the reflection of the actual band. (b) The induced solutocapillary flow transports $CO_2$ rich layer away from the source and into the bulk. }
\label{fig:PLIF_whole}
\end{figure}

The PLIF technique can be used to visualize the solutocapillary flow induced at and near the water surface, including its start-up behavior, by monitoring the $CO_2$ concentration profile as a function of time. As discussed above, a thin layer of $CO_2$ rich region, about $0.5 mm$ thick, is formed near the water surface by releasing $CO_2$ into the chamber at the beginning of the experiment (see figure \ref{fig:PLIF_whole}). Since the rate of diffusion of $CO_2$ is small, the layer diffuses slowly into the bulk, but it is readily convected by the flow. Thus, the time evolution of the $CO_2$ concentration profile can be used to visualize the induced solutocapillary flow. The $CO_2$ rich layer in this figure is transported by the solutocapillary flow induced by a freshwater source present in the middle of the photograph. The figure shows that the flow is symmetric and that it induces vortices in the near surface layer which cause $CO_2$ rich surface layer of water to mix with the water below. After the surface flow changes direction inwards, the radially outward transport of $CO_2$ rich surface layer stops.

\begin{figure}
  \centerline{\includegraphics{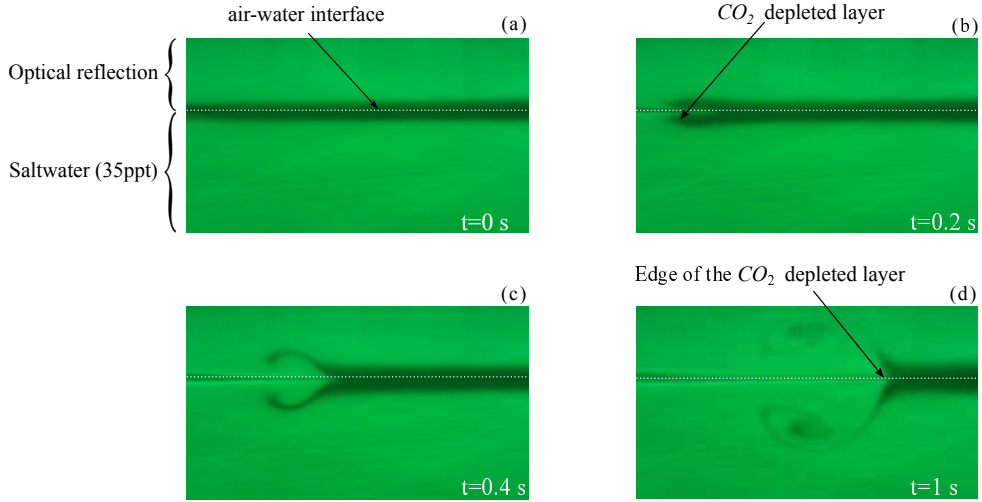}}
  \caption{Magnified PLIF images of CO2 concentration profiles at four different times. The source needle is present on the left side of the image. White dotted lines are added to distinguish the optical reflection from the bulk (a) Shortly after $CO_2$ was injected in the chamber. (b)-(d) The solutocapillary flow transported $CO_2$ layer away from the source}
\label{fig:PLIF_zoom}
\end{figure}
\begin{figure}
  \centerline{\includegraphics{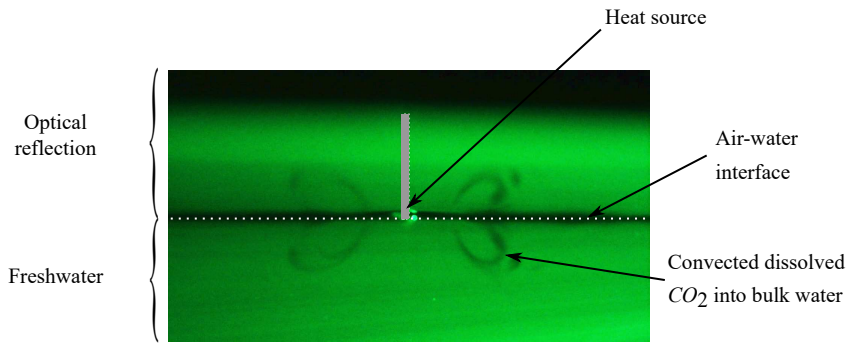}}
  \caption{A PLIF image of $CO_2$ transport by the thermocapillary flow induced by a cylindrical heat source present on the surface. The half angle of the conical inflow region is around $60^{\circ}$ and the $CO_2$ rich layer moves inwards after a few millimeters}
\label{fig:themo_PLIF}
\end{figure}
Figure \ref{fig:PLIF_zoom} shows the magnified views of the $CO_2$ concentration profiles as a layer of $CO_2$ is transported by the solutocapillary flow induced by a freshwater source present on the surface, which is out of view and about one centimeter to the left of the photograph (not visible in the photograph). The induced flow on the surface is towards the right and below the surface it is towards the needle, and so the $CO_2$ rich dark layer of water on the surface is swept to the right (away from the needle) and its place is taken by the water from below which has smaller $CO_2$ concentration (hence colorless). Furthermore, as the water velocity is maximal at the surface and decreases sharply with increasing distance from the surface, the part of the $CO_2$ rich layer at and near the surface is convected to a greater distance than the part slightly below the surface which creates a colorless band near the surface in which the $CO_2$ concentration is smaller (and also the salt concentration). As result, the left end of the layer in figure \ref{fig:PLIF_zoom}(b) looks like a horseshoe (however, note that the upper part of the shoe in the photograph is the reflection of the lower part.) The thickness of this colorless band is $\approx 0.5 mm $ which is of the order as the momentum boundary layer thickness. In contrast, the thermocapillary flow on the surface induced by a hot rod persisted for only a few millimeters and then it changed direction inwards (see figure \ref{fig:themo_PLIF}). As a result, in this case $CO_2$ was transported in a relatively smaller region near the heat source and the half cone angle for inflow was around $60^{\circ}$.

The strength of solutocapillary flow weakens with increasing distance from the needle and the thickness of $CO_2$ rich layer below the colorless band increases, as it is pushed downwards away from the interface. As noted earlier, for the source flow rates considered in this study, the radially outward flow on the surface changed direction into the water at a distance of a few centimeters from the needle depending on the source flow rate. Beyond this distance, the streak lines of $CO_2$ leave the surface at an angle into the water body. The latter is consistent with the velocity distribution of figure \ref{fig:piv_vel_3fr}. Also notice that the $CO_2$ concentration profile on the surface does not change beyond this distance as the convection transport is not present. This also suggests that the salt concentration is approximately constant beyond this point which is consistent with the fact that there is no solutocapillary flow in this region (in agreement with the measurements reported in \citet{Roche2014} which show that the surfactant concentration is approximately constant after the surface flow changes direction into the waterbody). The water moving downward away from the surface carries $CO_2$ rich water, which enhances the mixing of the surface water into the bulk. The transient PLIF concentration profiles also indicate that in the region where the flow is approximately tangential to the surface, the solute transport away from the surface is small since only when the flow leaves the surface at an angle that the $CO_2$ and solute mix in the water below.

\section{Summary and discussion}

In summary, the focus of this paper is on the solutocapillary flow induced on and below the surface by a freshwater point source present on the surface of a saltwater body. We find that the velocity field measured using the PIV technique is in good agreement with the analytic solution, except near the source needle and after few centimeters away from the needle where the surface flow changes its direction into the bulk.

Specifically, for both the analytic solution and experiments, $u_r <0$ in a conical region with its vertex at the source and $u_r >0$ outside the conical region. The velocity magnitude is locally minimal along the cone boundary. For the analytic solution, the velocity components assume positive and negative values such that the net liquid flow rate from the source is zero, and become singular as $r$ approaches zero. The flow induced by a solute source is thus analogous to that induced by a doublet consisting of a source-sink pair, except that the solutocapillary flow is limited to the half space below the water surface. The flow for a positive solute source is qualitatively similar except that the flow direction is reversed.

The half cone angle for the induced flow varies with the solute inflow rate, especially for a negative solute source as the critical value of $\theta$ at which the direction of radial flow reverses increases with the solute inflow rate. At small solute inflow rates, the angle is about $60^{\circ}$, and increases with increasing solute inflow rate to values greater than $80^{\circ}$. Under the latter conditions, the velocity and concentration boundary layers develop near the surface as the outflow is restricted to a thin axisymmetric wedge shaped boundary layer region. Again, because of the separable form of the solution, the concentration and momentum boundary layers are given by: $\theta=\upi/2- \alpha_c$ and $\theta=\upi/2-\alpha_m$, respectively. Here, the constants $\alpha_c$ and $\alpha_m$ are small angles which decrease with increasing $Sc$ and $Q_c$. The concentration and momentum boundary layer thicknesses, given by $\delta_c=r \sin{\alpha_c}$ and $\delta_m= r\sin{\alpha_m}$, vary linearly with $r$, and their ratio $\delta_m/\delta_c$   is independent of $r$. For the solute flow rates considered, the solute boundary layer thickness $(\approx 0.05 mm)$ is an order of magnitude smaller than the momentum boundary layer $(\approx 1 mm)$.

The Schmidt number $Sc$ plays an important role in determining the strength of solutocapillary flow and the distance for which the flow persists. This happens because when $Sc$ is large the solute gradient on the surface diminishes slowly with the distance from the source and so it continues to drive the flow for a longer distance. According to the analytic solution, when $Sc$ is increased from $447$ to $1000$ by reducing the solute diffusion coefficient by a factor of about two while keeping all other parameters fixed, the momentum boundary layer thickness at a distance of $1 cm$ decreases by $\approx10\%$ and the solute boundary layer thickness decreases by $\approx40\%$, and the surface tension gradient and velocity increase by $\approx31\%$. This is due to the fact that the solute concentration gradient gives rise to the force that drives the flow and so if $D$ is reduced, the flow is stronger and persists longer.

In contrast, when $Sc=7$ the solute boundary layer thickness is only about two times smaller than that of the momentum boundary layer. This is the case for thermocapillary flows for which the Prandtl number, which plays the same role as the Schmidt number for the solutocapillary flows, is around $7$. Thus, although the velocity fields for these two surface tension gradient driven flows are qualitatively similar, there are some important differences as the half cone angle for thermocapillary flows is closer to $60^{\circ}$ and the surface flow persists for a relatively shorter distance.

The flow slows down with increasing distance from the source and after a few centimeters the surface flow changes direction into the waterbody. The flow beyond this distance does not match with the analytic solution. A similar change in the flow direction happens in other surface tension driven flows at a distance which depends on the Schmidt and Marangoni numbers. This change in the flow direction enhances mixing as it convects the surface water with dissolved gases into the waterbody. The PLIF data also shows that the solute is not convected beyond this distance and so the surface tension gradient is negligible. This was also found to be the case for the surfactant induced Marangoni flows for which the surfactant concentration was constant beyond the distance after which the surface flow direction changes inwards \citep{Roche2014}

Analysis of the experimental and analytic results shows that the surface flow changes direction at a distance where the solute induced surface tension gradient {\it{reduces}} to the same order of magnitude as that for the Raleigh-Bernard convection due to evaporative cooling. Also, the expression for the distance at which the surface flow changes direction, obtained by setting the solute and RB convection induced surface tension gradients equal, correctly estimates the distance over which the surface flow occurs as a function of the source flow rate.

\appendix
\section{}\label{app}
\begin{figure}
  \centerline{\includegraphics{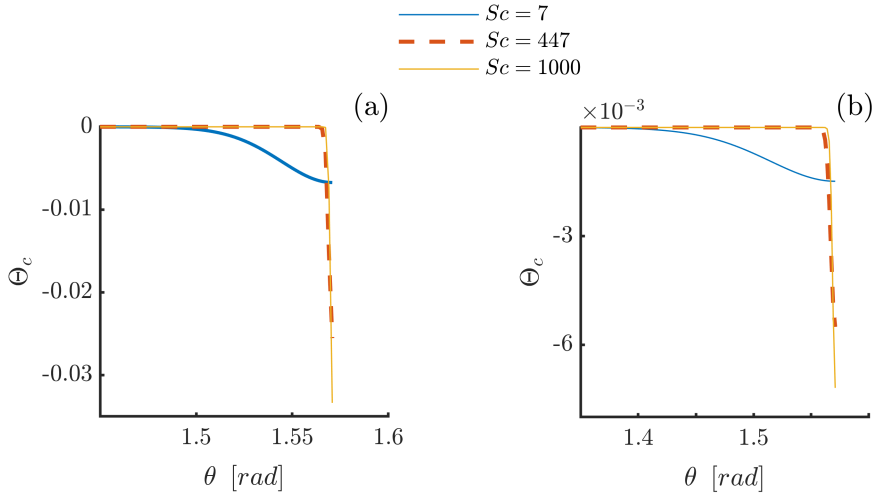}}
  \caption {$\Theta_c$ as a function of $\theta$ for three different values of $Sc = 7, 447$ and $1000$.(a) $Q_c = - 2.33\times 10^{-7} kg/s$, $Ma = 1.1\times10^{5}, 7.0\times10^{6}, 1.57\times10^{7}$. (b)  $Q_c = - 2.33\times 10^{-8} kg/s,  Ma = 1.1\times10^{4}, 7.0\times10^{5}, 1.57\times10^{6}$. }
\label{fig:thetaC_theta}
\end{figure}

\begin{figure}
  \centerline{\includegraphics{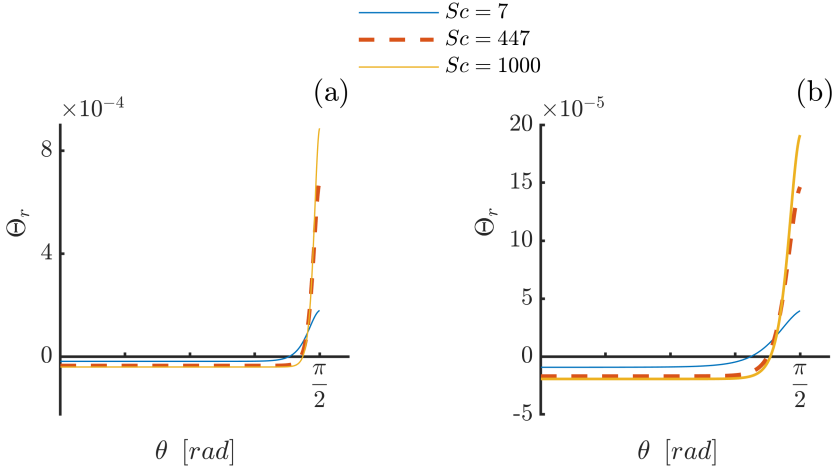}}
  \caption{$\Theta_r$ as a function of $\theta$ for three different values of $Sc = 7, 447$ and $1000$.(a) $Q_c = - 2.33\times 10^{-7} kg/s$, $Ma = 1.1\times10^{5}, 7.0\times10^{6}, 1.57\times10^{7}$. (b)  $Q_c = - 2.33\times 10^{-8} kg/s,  Ma = 1.1\times10^{4}, 7.0\times10^{5}, 1.57\times10^{6}$.}
\label{fig:Theta_r_theta}
\end{figure}

\begin{figure}
  \centerline{\includegraphics{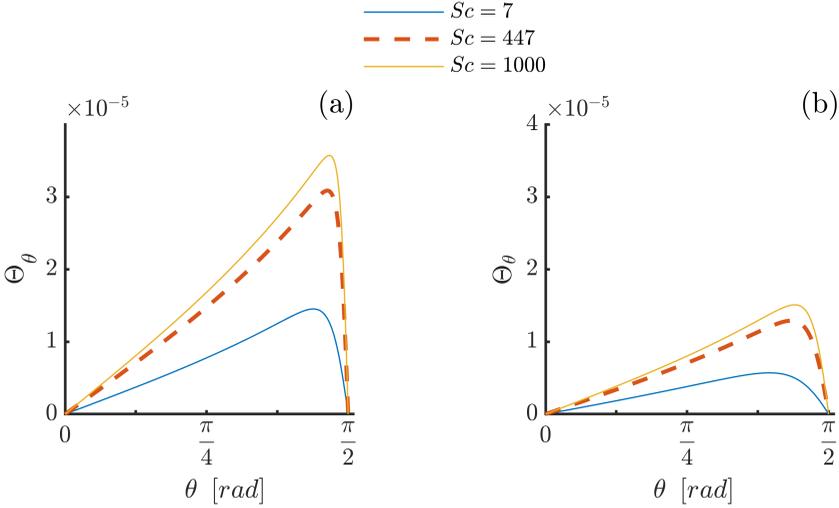}}
  \caption{$\Theta_\theta$ as a function of $\theta$ for three different values of $Sc = 7, 447$ and $1000$.(a) $Q_c = - 2.33\times 10^{-7} kg/s$, $Ma = 1.1\times10^{5}, 7.0\times10^{6}, 1.57\times10^{7}$. (b)  $Q_c = - 2.33\times 10^{-8} kg/s,  Ma = 1.1\times10^{4}, 7.0\times10^{5}, 1.57\times10^{6}$.}
\label{fig:Theta_theta_theta}
\end{figure}

In this section, we discuss the analytic solution given by equations \ref{ur}-\ref{conc}, including its dependence on the parameter values. In figure \ref{fig:thetaC_theta}, $\Theta_c$ is shown as a function of $\theta$ for three different values of $Sc$, with $Q_c  = - 2.33\times 10^{-7} kg/s$ in figure \ref{fig:thetaC_theta}(a) and $Q_c = - 2.33\times 10^{-8} kg/s $ in figure \ref{fig:thetaC_theta}(b). Figures \ref{fig:Theta_r_theta} and \ref{fig:Theta_theta_theta} show $\Theta_{r}$ and  $\Theta_{\theta}$ as a function of $\theta$ for the same parameter values. The value of $Sc$ in these figures is increased by reducing $D$ while keeping the remaining parameters fixed. Figure \ref{fig:thetaC_theta} shows that near the liquid surface, i.e., $\theta$ near $\frac{\upi}{2}$, the concentration varies slowly with $\theta$ for $Sc = 7$. However, for $Sc=447$ and $1000$, a boundary layer of the solute concentration develops in which the solute concentration is sharply reduced while the concentration away from the surface remains approximately zero. Notice that the concentration at the surface is negative because of the presence of a negative solute source which reduces the concentration from the mean bulk concentration that is assumed to be zero. Also, notice that the concentration at the solute source is not specified in the analytic solution, and that only the solute in- or out-flow rate is specified. Since the concentration far away from the source is assumed to be zero, for a positive solute source, $c > 0$ near the source and approaches infinity, as $r\rightarrow0$. For a negative source (sink), $c < 0$ near the source and approaches minus infinity, as $r\rightarrow0$.\\
The maximum value of ${|\Theta}_c|$ increases and the thickness of the boundary layer decreases with increasing Sc. Thus, the gradient of solute concentration at the surface increases with increasing $Sc$ which makes the liquid velocity near the surface larger and $\Theta_{r}$ develops a boundary layer near the surface while $\Theta_{\theta}$ remains small. Notice that the thickness of the solute concentration boundary layer is smaller than that of the velocity boundary layer. For both boundary layers the thickness is defined to be the distance from the surface at which the value becomes $1\%$ of the value at the surface. Also, due to the separable nature of the analytic solution, the thickness increase linearly with the distance from the source. 

\begin{figure}
  \centerline{\includegraphics{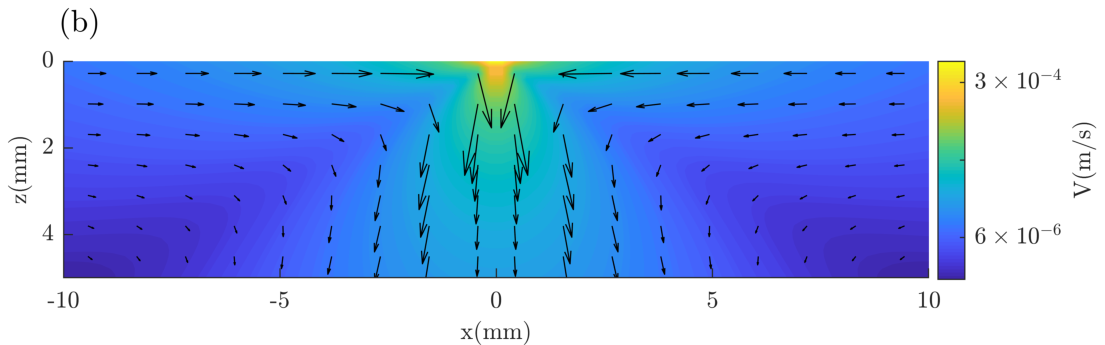}}
  \caption{The velocity field on a vertical plane passing through the source for $Sc = 447$. To improve clarity, the arrows near the origin (source) where the velocity is relatively large are not plotted, and the color bar is in logarithmic scale. Positive source with $k = -0.5$, $Q_c = 2.33\times10^{-7} kg.s (Ma=7.00\times 10^{-7}).$}
\label{fig:vel_field_anal}
\end{figure}

\begin{figure}
  \centerline{\includegraphics{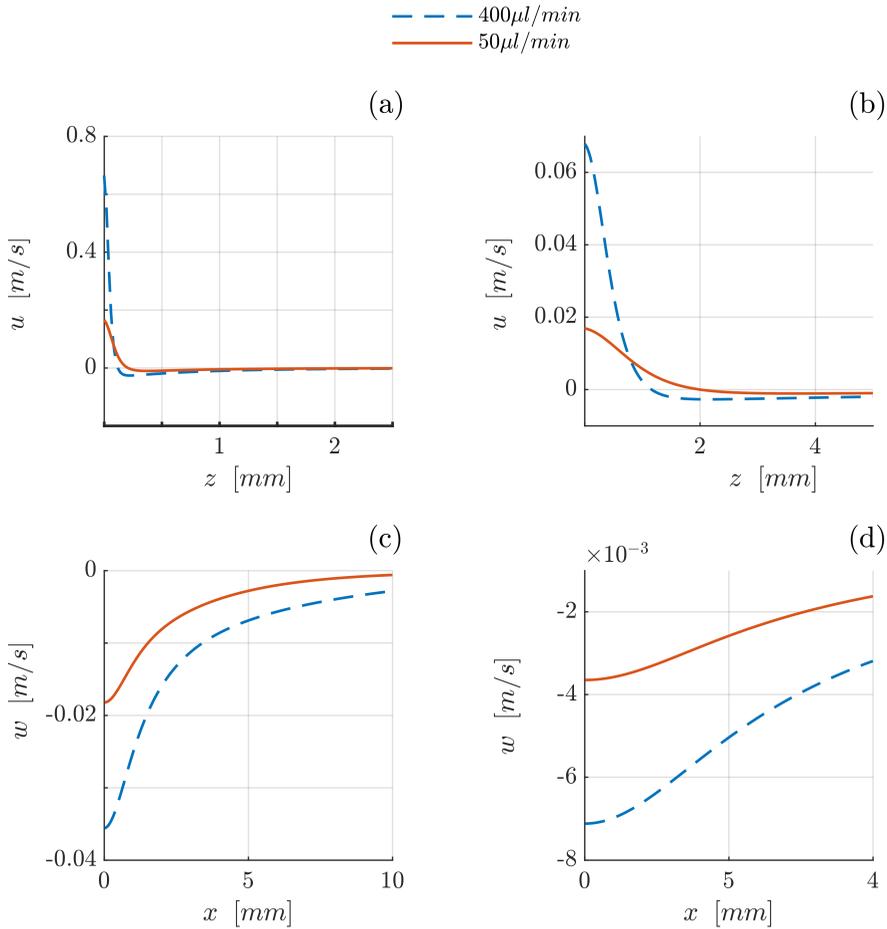}}
  \caption{The velocity as a function of $x$ and $z$ and for two negative source strengths, $Q_c  = -2.33\times10^{-7} kg/s$, $-2.91\times10^{-8} kg/s$ ($Ma = 7.0\times10^6, 8.75\times10^5$) and $Sc = 447$. These values correspond to the fresh water flow rates of $50 \mu l/min$ and $400 \mu l/min$, respectively, in our experiments. 
  (a) $u$ as a function of $z$ for $x= 1 mm$ 
  (b) $u$ as a function of $z$ for $x= 10.0 mm$ 
  (c) $w$ as a function of $x$ for $z = 1 mm$
  (d) $w$ as a function of $x$ for $z = 5 mm$.  }
\label{fig:u_w_x_z}
\end{figure}

\begin{figure}
  \centerline{\includegraphics{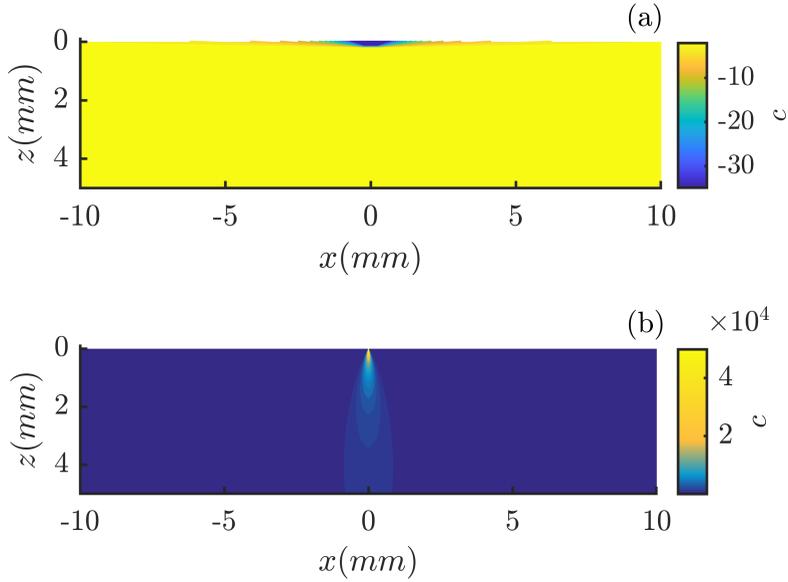}}
  \caption{Isovalues of solute concentration $c$ on a vertical plane passing through the source for $Sc = 447$, (a) Negative source, $K = 1525.8$ and $Q_c = -2.33\times10^{-7} kg/s$ ($Ma = 7.0\times10^7$). The solute concentration is reduced near the source and the flow on the surface is away from the source, and so the region of small $c$ emanates radially outward along the surface. (b) Positive source, $K = -0.5$ and $Q_c = 2.33 \times10^{-7} kg/s$ $(Ma = 7.0\times10^7)$. The solute concentration is larger near the source and the flow near the source is in the downward direction, and so the region of large $c$ emanates downward from the source.}
\label{fig:concentration_contour}
\end{figure}

\begin{figure}
  \centerline{\includegraphics{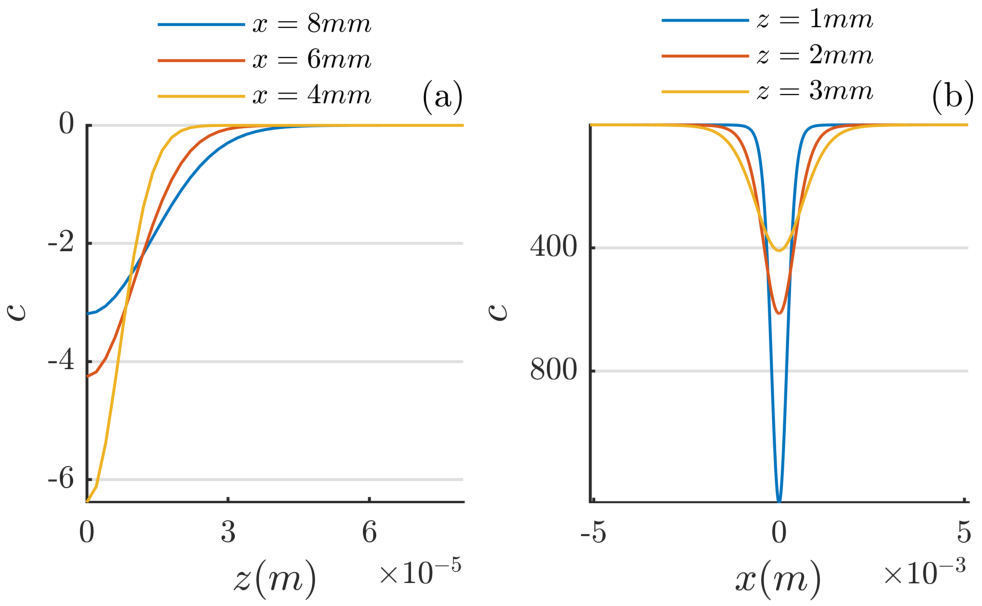}}
  \caption{Solute concentration $(c)$ as a function of $x$ and $z$ on a vertical plane passing through the source for $Sc = 447$. (a) For a negative source $(K =1525.8)$, $c$ is shown as a function of $z$ for fixed $x$-values. $Q_c = 2.33\times10^{-7} kg/s$ and $Ma = 7.0\times10^7$ (b) Positive source, $K = -0.5$, $c$ is shown as a function of $x$ for fixed $z$-values. $Q_c = 2.33 \times10^{-7} kg/s$ and $Ma = 7.0\times10^7$.}
\label{fig:concentratoin_depth}
\end{figure}

We next describe the velocity and solute concentration fields in the physical space. In figures \ref{fig:vel_doublet}(c) and \ref{fig:vel_field_anal} the velocity field is shown for a solute sink $(Q_c<0$) and a solute source $(Q_c>0)$, and in figure \ref{fig:u_w_x_z} the velocity components are plotted for fixed $x$ or $z$ values. The solute concentrations are shown is figures \ref{fig:concentration_contour} and \ref{fig:concentratoin_depth}. Figures \ref{fig:vel_doublet}(c) and \ref{fig:concentration_contour}(a) show that the out-flow of solute (due to the solute sink) decreases the local solute concentration which decreases the surface tension near the sink giving rise to a flow on the surface away from the sink. This in turn causes a decrease in the solute concentration on and near the surface along the outflow direction. The opposite is true for a solute source (see figures \ref{fig:vel_field_anal} and \ref{fig:concentration_contour}(b)). The inflow of solute increases the local salt concentration and this in turn increases the local surface tension which drives a flow on the surface towards the source and below the surface away from the source. The flow away from the source convects liquid with higher solute concentration downwards. 

The solute transport is convection dominated, as can be seen from the solute iso-concentration levels of figure \ref{fig:concentration_contour}. Therefore, since for a negative source the solute concentration near the source is smaller and on the surface the liquid moves radially outward from the source, the liquid with low solute concentration is convected outward on the surface. This results in the formation of a solute concentration boundary layer, i.e., a thin layer in which the solute concentration varies sharply in the direction normal to the flow direction and is considerably depleted (figure \ref{fig:concentration_contour}). In figure \ref{fig:concentratoin_depth}, at a distance of $4 mm$ from a negative source, the solute boundary layer thickness is only a fraction of one millimeter. The thickness of the solute boundary layer increases with increasing distance from the source due to diffusion (see figure \ref{fig:concentratoin_depth}). This is also the case for the velocity boundary layer thickness which increases with increasing distance from the source (see figure \ref{fig:u_w_x_z}). Also, as noted above, the thickness of velocity boundary layer is larger than that of the concentration boundary layer.


For a positive source, the solute concentration near the source is larger and the flow on the surface is towards the source and below the source it is away from the source and so the solute rich liquid near the source is convected downward resulting in the iso-concentration levels as shown in figure \ref{fig:concentration_contour}(b). This results in a boundary layer below the source, in the outflow direction, where the velocity is larger. Furthermore, as figure \ref{fig:u_w_x_z} shows, when $|Q_c|$ is increased, the magnitude of velocity increases and the thickness of the velocity boundary layer decreases as the relative importance of convection increases. 

\clearpage

\bibliographystyle{jfm}
\bibliography{Soluto_capillary_flow.bib}

\end{document}